\documentclass[a4paper,10pt,twoside]{cpc-hepnp}

\usepackage{multicol}
\usepackage{graphicx}
\usepackage{booktabs}
\usepackage{amssymb,bm,mathrsfs,bbm,amscd}
\usepackage[tbtags]{amsmath}
\usepackage{lastpage}
\usepackage{CJK}
\usepackage{color}
\usepackage{epstopdf}
\usepackage{epsfig}
\usepackage{overpic}
\usepackage{flushend}
\usepackage{float}

\begin{document}
\begin{CJK*}{GBK}{song}
\title{QCD phase diagram at finite isospin chemical potential and temperature in an IR-improved soft-wall AdS/QCD model}

\author{%
Xuanmin Cao$^{1)}$
\email{caoxm@jnu.edu.cn} 
\quad Hui Liu $^{2)}$
\email{tliuhui@jnu.edu.cn}
\quad Danning Li $^{3)}$
 \email{lidanning@jnu.edu.cn} 
\quad Guanning Ou $^{4)}$
 \email{ouguanning@outlook.com}
 }
 \maketitle
\address{%
Department of Physics and Siyuan Laboratory,  Jinan University, Guangzhou 510632, Peoples Republic of China
}
\end{CJK*}

\begin{abstract}
We study the phase transition between pion condensed phase and normal phase, as well as chiral phase transition in a two flavor($\mathcal{N}_f=2$) IR- improved soft-wall AdS/QCD model at finite isospin chemical potential $\mu_I$ and temperature $T$. By self-consistently solving the equations of motion, we obtain the phase diagram in the plane of  $\mu_I$ and $T$.  The pion condensation appears together with a massless Nambu-Goldstone boson $m_{\pi_1}(T_c, \mu_I^c)=0$, which is very likely to be a second-order phase transition with mean-field critical exponents in small $\mu_I$ region. When $T=0$, the critical isospin chemical potential approximates to vacuum pion mass $\mu_I^c \approx m_0$. The pion condensed phase exists in an arched area and the boundary of the chiral crossover intersects the pion condensed phase at a tri-critical point. Qualitatively, the results are in good agreement with previous studies from Lattice simulations and model calculations.
\end{abstract}

\date
\begin{multicols}{2}

\section{Introduction}\label{sec:intro}
Quantum chromodynamics (QCD) at finite isospin chemical potential $\mu_I$ has attracted more and more attentions in the study of chiral symmetry breaking ($\chi$SB) as well as the color confinement mechanism of strong interaction~\cite{QGP}. In the heavy-ion collision, it is still unclear how the medium evolves through the QCD phase transition with finite baryon chemical potential $\mu_B$ and isospin chemical potential $\mu_I$ at finite temperature\cite{isospinHIC}.  In astrophysics, the neutron star possesses large imbalance between the density of neutrons and protons at very low temperature, which also attracts people's attention on QCD at finite $\mu_I$~\cite{neutronstars}.

A number of theoretical methods are available to study QCD phase transitions at finite $\mu_I$, including the chiral perturbation theory ($\chi$PT)~\cite{D.T.Son1,G.Baym,D.B.Kaplan, C.Michael, A.Mammarella,M.Loewe1,Carignano2017,L.Lepori, P.Adhikari}, the Nambu-Jona-Lasinio (NJL) models \cite{A.Barducci,D.Toublan,A.Barducci1,A.Barducci2,P.F.Zhuang,P.F.Zhuang3,P.F.Zhuang2,chao2018charged,zhang2018mesonic}, the quark-meson models \cite{PhysRevD.95.036017,PhysRevD.98.074016,PhysRevD.97.076005,PhysRevD.96.014006}, the linear sigma models~\cite{PhysRevD.89.016004, PhysRevD.94.056012}, the random matrix models \cite{PhysRevD.68.014009,  PhysRevD.72.015007} and the perturbation QCD (pQCD) \cite{PhysRevD.93.085030, Haque2014}. These theoretical methods give us comparable results of the meson condensed phase with non-zero $\mu_I$. Lattice QCD (LQCD) is usually considered as one of the most powerful first-principle calculations to explore  non-perturbative QCD at finite temperature. However, the notorious negative sign problem of fermion determinant makes it very difficult to study the system at finite baryon chemical potential $\mu_B$~\cite{I.Barbour, J.B.Kogut1}. Fortunately, there is no such sign problem at finite $\mu_I$ \cite{M.Alford}, and LQCD is extensively used in non-zero isospin systems\cite{J.B.Kogut2,J.B.Kogut3,N.Yusuke, J.B.Kogut,S.R.Beane,W.Detmold,B.Brandt, B.Brandt1, B.Brandt2}. These studies provide numerical evidence for the proposed meson condensation phase at finite $\mu_I$. The $\mu_I$ axis can roughly be divided  into two parts by the point $2 m_{0}$, with $m_0$ the vacuum pion mass. In the region of $0<\mu_I<2 m_{0}$, the lattice results are consistent with other theoretical analysis and model calculations. For example, it is shown that the location of zero temperature critical isospin chemical potential is $\mu_I^c=m_0$, and it is a second order transition from the normal phase to pion condensation at low temperature. At large $\mu_I$ region, LQCD is hard to control due to the lattice saturation effects \cite{J.B.Kogut}. However, we notice that both in LQCD study\cite{J.B.Kogut} and NJL study\cite{P.F.Zhuang}, the large $\mu_I$ transition line in $T-\mu_I$ plane would bend towards the $\mu_I$ axis, which is completely different from others which shown tendencies to increase or saturate. For a review of meson condensation with finite $\mu_I$, please refer to Ref.~\cite{M.Mannarelli}.

The experimental data and theoretical predictions suggest that the QGP  is probably strong coupled~\cite{ADAMS2005102,PhysRevC.87.014902,PhysRevLett.105.252302,SHURYAK2004273,PhysRevLett.87.081601,PhysRevLett.93.090602,PhysRevLett.94.111601}.
To handle the tough strong coupling problems, the holography method~\cite{J.Maldacena,E.Witten, S.Gubser, S.Gubser1} has been widely applied in many fields, such as nuclear physics~\cite{policastro2002ads,karch2002adding,Natsuume2015} and condensed matter physics \cite{Sachdev2011,Herzog2009}. To mimic QCD physics, in the bottom-up framework, the hard-wall AdS/QCD model Refs.~\cite{D.T.Son0} and soft-wall AdS/ QCD model Ref.~\cite{D.T.Son} have been constructed.  In the hard-wall model, the chiral symmetry breaking can be well described. However, the linear Regge behavior of the hadron spectrum is not depicted in the model. The original soft-wall model can describe the correct Regge behavior of the meson spectrum by introducing an infrared (IR) suppressed dilaton term. Also, it is also quite natural to introduce chemical potential through the 5D gauge field, which is dual to the 4D conserved current. Thus, it produces a good start point to study physics related to linear confinement and chiral symmetry breaking, both at finite temperature and at finite densities. We also noticed a series of work \cite{G_rsoy_2008,G_rsoy_20082,PhysRevD.81.115004,Iatrakis2010,Jarvinen2012,Alho2013,Alho2014,Jarvinen2015,CASERO200798}, in which holographic QCD is constructed with more stringy ingredients. The properties of QCD thermodynamics and chiral phase transition in the Veneziano limit have been investigated in detail.

In this paper, we will try to investigate QCD phase diagram, more concretely, the properties of the pion condensation and chiral condensation, at finite $\mu_I$ and $T$ in the framework of soft wall model. The relevant issues have been studied in the well-known Sakai-Sugimoto model in the probe approximation with external magnetic field~\cite{Rebhan2009,Aharony2008}. Actually, in the hard-wall AdS/ QCD model,
the pion condensation has been studied by introducing a baryonic charge in the IR boundary at zero temperature~\cite{enhancement1, enhancement2, Albrecht2010}. It has been shown that the phase transition, between normal phase and pion condensation phase, is of second-order with mean-field critical exponents and happens at $\mu_I^c=m_0$.
Efforts have also been made in the soft-wall framework at finite temperature \cite{D.Li}. It shows that the phase transition is of first-order with two both left and right critical $\mu_I$ at a particular temperature, which are quite different from the hard wall results. This modified soft wall model is constraint from chiral phase transition and the meson spectrum of this model is still unclear. Therefore, it is still interesting to investigate the phase transition in a model describing the experimental meson spectrum. We note that an IR-improved soft-wall AdS/QCD model, proposed in \cite{Z.Fang}, can generate both the chiral spontaneous breaking and  the  linear Regge behavior of the hadron spectrum. Especially, the vacuum pion mass $m_0=139.6$ MeV is well described. So that we would like to adopt such an IR-improved soft-wall AdS/QCD model in this paper and we expect to compare our results with those results of  LQCD and analytical theories or models.

This paper is organized as follows. In section~{\ref{1:model1}}, we review the IR-improved soft-wall model and introduce $\mu_I$ to this model, then derive the effective Lagrangian. In section~{\ref{QCDphase}}, we obtain the equations of motion of scalar, pseudo-scalar and axial-vector fields, then study the chiral transition and pion condensation at finite $\mu_I$ and $T$. In section~{\ref{phasediagram}}, we give a complete phase diagram of chiral condensation and pion condensation in the $\mu_I-T$ plane. At last, we give a conclusion and discussion in section~{\ref{conclusion}}.

\section{The IR-improved soft-wall AdS/QCD model with finite $\mu_I$}\label{1:model1}
The IR-improved soft-wall AdS/QCD model~\cite{Z.Fang} is constructed in the bottom up framework~\cite{D.T.Son0,D.T.Son} with a quartic term of bulk scalar and a modified 5D conformal mass of the bulk scalar field.
The background spacetime of this soft-wall model is the following $\rm{AdS}_5$ spacetime metric,
\begin{equation}
d s^2=e^{2 A(z)}(\eta_{\mu\nu} d x^\mu dx^\nu - d z^2) ,
\end{equation}
where $\eta^{\mu\nu}={\rm{diag}}\{+1,-1,-1,-1\}$, $z$ is the holographic radial coordinate and $A(z)=-{\rm{ln}}(z/L)$ with $L$ the AdS curvature radius which will be set to unit for simplicity in the following calculation.

On top of this background geometry, the soft-wall AdS/QCD model with $\mathcal{N}_f=2$ is constructed with $SU(2)_L\times SU(2)_R$ gauge symmetry. The meson sector of the 5D action can be written as
\begin{eqnarray}
S_M&=&\int d^5 x \sqrt{g} e^{-\Phi (z)} {\rm{Tr}} \{ |D X|^2-m_5^2(z)|X|^2-\lambda|X|^4\nonumber\\
& &-\frac{1}{4g_5^2}(F_L^2+F_R^2)\},
\end{eqnarray}
where $g$ is the determinant of the metric $g_{MN}$,  $\Phi(z)=\mu_g^2 z^2$ is the dilaton profile with $\mu_g$ a constant
mass scale necessary for the Regge behavior of meson spectrum\cite{D.T.Son}. $g_5$ is the gauge coupling which can be determined by comparing the large momentum expansion of correlator of vector current $J_\mu^a=\bar{q}\gamma_\mu t^a q$ in both AdS/QCD and perturbative QCD\cite{D.T.Son0} where $t^a$ ($a=1,2,3$) is the generators of $\rm{SU}(2)$.
In general, the field $X$,  which is a complex $2\times 2$ matrix valued bulk scalar filed, can be decomposed into the pseudo-scalar meson field $\pi(x,z)=\pi^a(x,z) t^a$ and the scalar meson field $S(x,z)=S^a t^a$ in the form of
\begin{equation}
X=(\chi t^0+S) e^{2 i\pi+i\eta},
\end{equation}
where $t^0=\rm{I}_2/2$ and $\chi(z)$ is related to the vacuum expectation value (VEV) of bulk scalar field $X$ by $\langle{X}\rangle= \rm{I}_2\chi/2\ $ with $\rm{I}_2$ the $2\times 2$ identity matrix.

In order to get a consistent description of both meson spectrum and chiral symmetry spontaneously breaking, the 5D mass $m_5^2(z)$, which relate to the quark mass anomalous dimension, can be modified by comparing the ultraviolet (UV) boundary and infrared (IR) boundary expression of the equation of motion (EOM) of the VEV of bulk scalar field~\cite{Z.Fang}, expressed as
\begin{equation}
m_5^2(z)=-3-\mu_c z^2,
\end{equation}
where $\mu_c$ is a free parameter fixed by fitting meson spectra. The leading constant term $-3$ can be determined from the AdS/CFT dictionary $m_5^2(z)=(\Delta-p)(\Delta+p-4)$ by taking $p=0$ and $\Delta=3$, which is the dimension of the dual operator $\bar{q}_R q_L$\cite{D.T.Son0}.

The covariant derivative $D^M$ and chiral gauge field strength $F_{L/R}^{MN}$ are defined as
\begin{subequations}
\begin{eqnarray}
D^M&=&\partial ^M X-i A_L^M X+i X A_R^M, \\
F_{L/R}^{MN}&=&\partial^MA_{L/R}^N-\partial^NA_{L/R}^M-i[A_{L/R}^M,A_{L/R}^N],
\end{eqnarray}
\end{subequations}
where $A_{L/R}^M=A_{L/R}^{a,M}t_{L/R}^a$, and the chiral gauge fields $A_{L/R}^M$ are dual to relevant QCD operators at the boundary by the AdS/QCD dictionary\cite{D.T.Son,D.T.Son0}.

For convenience, one can redefine the chiral gauge fields into the vector gauge field and the axial-vector gauge field,
\begin{eqnarray}
V^M=\frac{A_{L}^M+A_R^M}{2},\ \ \ A^M=\frac{A_{L}^M-A_R^M}{2},
\end{eqnarray}
then one has the covariant derivative and   transformed chiral gauge field strength as,
\begin{subequations}
\begin{eqnarray}
D_M X&=& \partial_M X-i[V_M,X]-i\{A_M,X\}
,\\
F_A^{MN}&=&\frac{1}{2}(F_L^{MN}-F_R^{MN})\nonumber\\
&=&\partial^MA^N-\partial^NA^M-i[V^M,A^N]-i[A^M,V^N],\nonumber\\
& &\\
F_V^{MN}&=&\frac{1}{2}(F_L^{MN}+F_R^{MN})\nonumber\\
&=&\partial^MV^N-\partial^NV^M-i[V^M,V^N]-i[A^M,A^N].\nonumber\\
\end{eqnarray}
\end{subequations}

Take the temperature and isospin chemical potential effects into account, instead of the pure $\rm AdS_5$ space, the AdS/ Reissner-Nordstrom (AdS/RN) black hole should be considered as the bulk background, then the metric ansatz
\begin{equation}
ds^2=e^{2A(z)}\bigg(f(z)dt^2-dx_i dx^i-\frac{dz^2}{f(z)}\bigg).
\end{equation}
For simplicity, we take the following metric solution with finite $\mu_I$,
\begin{subequations}
\label{domuI}
\begin{eqnarray}
A(z)&=&-{\rm{ln}} (z),\\
f(z)&=&1-(1+\gamma \mu_I^2z_h^2)\frac{z^4}{z_h^4}+\gamma\mu_I^2\frac{z^6}{z_h^4},\\
v&\equiv &V_0^3(z)=\mu_I\bigg (1-\frac{z^2}{z_h^2}\bigg),
\end{eqnarray}
\end{subequations}
where $\gamma$ is related to the coupling of $V_0^3$ with gravity, which can be taken as a free parameter and we set $\gamma=1$. Different from Refs.~\cite{enhancement1,enhancement2}, we will take $V_0^3$ as a background field other than a dynamical field in the following discussion.
The temperature could be introduced if there is a horizon $z=z_h$ where $f(z)=0$. The temperature is related to $z_h$ by the formula
\begin{equation}\label{temperature}
T=\frac{1}{4\pi}\bigg|\frac{df(z)}{dz}\bigg|_{z=z_h}=\frac{2-\gamma \mu_I^2 z_h^2}{2\pi z_h},
\end{equation}
where we have used the solution in Eq.~(\ref{domuI}). By all these definitions, It is demanded that the outer horizon $z_h< \sqrt{2/\gamma \mu_I^2}$ to make sure positive temperature and $z_h= \sqrt{2/\gamma \mu_I^2}$ at $T=0$ other than $f\equiv 1$.

In the case of finite $\mu_I$ and $T$, one can check $S^a$, $\pi^a$ and $\eta$ vanish if there are no extra sources of the corresponding operators. The theory has the $ U_I(1)$ symmetry which is a subgroup of the isospin $ SU_I(2)$. As discussed in Ref.\cite{enhancement1}, using this $ U_I(1)$ symmetry, $V_i=R_i=0$, by choosing the special angle with vanishing condensation of the $\pi^2$ field and keep only the $\pi^1$ condensation, and the iso-triplet scalars do not condense, $S^a=0$. We let $\Pi\equiv\pi^1$, $V_0^1=V_0^2=0$, $A_0^3=\pi^3=0$, $\eta=0$ and $A_0^0=0$.
Under these assumptions, the effective Lagrangian in the 5D spaces becomes
\begin{eqnarray}\label{effectlagrangian}
\mathcal{L}_{eff}&=&
\frac{e^{A-\Phi} \left({a_1}'^2+{a_2}'^2\right)}{2 {g_5}^2}-\frac{1}{2} f e^{3 A-\Phi}\left(\chi^2 \Pi '^2+\chi '^2\right)-\nonumber\\
& &e^{5 A-\Phi} \left(\frac{1}{2} m_5^2 \chi^2+\frac{1}{8} \lambda  \chi^4\right)+\frac{\chi^2 e^{3 A-\Phi}}{2 f} \bigg({a_1}^2-\nonumber\\
& &{a_2} v \sin (2 \Pi)
+{a_2}^2 \cos ^2(\Pi)+v^2 \sin ^2(\Pi)\bigg),\nonumber\\
\end{eqnarray}
where $v$, $a_1$, and $a_2$ instead of $V_0^3$, $A_0^1$, and $A_0^2$, respectively, and $'$ respects for the derivative with respect to $z$.

\section{QCD phase transition}\label{QCDphase}
In this section, we will study the phase transitions among the pion condensation phase, normal chiral symmetry breaking phase ($\chi$SB), and normal chiral symmetry restored phase($\chi$SR).  Firstly,  we derive the equations of motion (EOMs) of scalar field, pseudo-scalar field and axial-vector fields  ($\chi(z)$, $\Pi(z)$, and $a_{1 (2)}(z)$). Secondly, we numerically solve the EOMs and extract the value of chiral condensate and pion condensate according to the holographic dictionary. Finally, we analysis the properties of phases and phase transitions in detail.
\subsection{Equations of Motion and boundary conditions}\label{sectionofeom}
In the equilibrium state, the system is homogeneous everywhere, therefore it is always good enough to neglect the fluctuation in the coordinate space.  By doing functional derivate of the action in Eq.~\eqref{effectlagrangian}, the corresponding EOMs of $\chi(z)$, $\Pi(z)$, and $a_{1 (2)}(z)$ are extracted as
\begin{subequations}\label{EOM}
\begin{eqnarray}\label{EOMa}
&&\frac{ \chi}{f^2}\left({a_2}^2  \cos ^2(\Pi)-{a_2}v \sin (2 \Pi)+v^2 \sin ^2(\Pi)-\Pi '^2 f^2\right)-\nonumber\\
&&\frac{e^{2 A} \chi}{f}\left(m_5^2+\frac{\lambda}{2}  \chi^2\right)+\chi ' \left(3 A'-\Phi '+\frac{f'}{f} \chi '\right)+\chi ''=0,\label{EOMa}\nonumber\\
&&
\end{eqnarray}
\begin{eqnarray}
&&\frac{v^2-{a_2}^2}{2 f^2} \sin (2\Pi )-\frac{{a_2} v \cos (2 \Pi )}{f^2}+\Pi ' \bigg(3 A'-\Phi '+\frac{f'}{f}+\nonumber\\
&& \frac{2 \chi '}{\chi }\bigg)+\Pi''=0,\label{EOMb}\\
&&-\frac{e^{2 A}  {g_5}^2 \chi ^2{a_1}}{f}+{a_1}' \left(A'-\Phi '\right)+{a_1}''=0,\label{EOMc}\\
&& \frac{\chi ^2 e^{2 A} {g_5}^2\left( v \sin (2 \Pi )-2{a_2} \cos ^2(\Pi )\right)}{2f}+{a_2}' \left(A'-\Phi '\right)+\nonumber\\
& &{a_2}''=0.\label{EOMd}
\end{eqnarray}
\end{subequations}
No explicit source exsits for the axial vector current and $a_1(z)$ does not appear in other equations, so that we can simply set $a_1(z)=0$ \cite{D.Li}.

Equations~\eqref{EOMa}, \eqref{EOMb}, and \eqref{EOMd},  are coupled nonlinear second order differential equations with multi-singular points, and they do not have exactly  analytical solutions. However, we can numerically solve them. Around the UV boundary ($z=0$), we get the expansion solutions of $\chi(z)$, $\Pi(z)$, and $a_2(z)$ as
\begin{subequations}\label{UVexpansion}
\begin{eqnarray}
\chi(z)&=& m_q \zeta  z+\frac{\sigma  z^3}{\zeta }+\frac{1}{2}   m_q\zeta  \left(-{\mu_c}^2+2 {\mu_g}^2+\frac{\lambda}{2} m_q^2\zeta ^2  \right)\nonumber\\
& &z^3 \ln (z)+\frac{1}{16}   m_q \zeta  \left(\mu_c^2-2 \mu_g^2- \frac{\lambda}{2} m_q^2 \zeta ^2  \right) \nonumber\\
& &\left(\mu_c^2-6 \mu_g^2-  \frac{3\lambda}{2}  m_q^2\zeta ^2\right)z^5 \ln (z)+\mathcal{O}(z^5),\\
\Pi(z)&=&\pi _1 z^2+\frac{1}{2} \pi _1 \left(\mu_c^2-2 \mu_g^2- \frac{\lambda}{2}  m_q^2\zeta ^2\right)z^4 \ln (z)+\nonumber\\
& &\frac{1}{8}  \bigg [a_{2c} \mu_I+\pi _1 \bigg(-\mu_c^2+6 \mu_g^2-\mu_I^2+\frac{\lambda}{2}  m_q^2 \zeta ^2- \nonumber\\
& &\frac{8 \sigma }{ m_q\zeta ^2}\bigg)\bigg ]z^4+\mathcal{O}(z^5),\\
a_2(z)&=&a_{2c} z^2+\frac{1}{8}  \left(a_{2c} g_5^2 m_q^2 \zeta ^2 +4 a_{2c} \mu_g^2- \pi_1 g_5^2 \mu_I m_q^2\zeta ^2 \right)\nonumber\\
& &z^4+\mathcal{O}(z^5),
\end{eqnarray}
\end{subequations}
where $m_q \zeta$, $\sigma/ \zeta$, $\pi_1$ and $a_{2c}$ are integral constants.
According to the holographic dictionary, we can identify coefficients, $m_q$, $\sigma$, and $\pi_1$, as quark mass, chiral condensate, and pion condensate, respectively. The normalization constant $\zeta=\sqrt{N_c}/2\pi$ is introduced to match the two point  function $\langle \bar{q} q(p), \bar{q} q(0)\rangle$ from holographic calculation and 4D calculation~\cite{A.Cherman}. Noticing that the external source is not considered in this work, so that the constant terms of $\Pi(z)$ and $a_2(z)$ are equal to zero. On the IR boundary (horizon with $z=z_h$), we have expansions solutions as
\begin{subequations}\label{IRexpansion}
\begin{eqnarray}
\chi(z)&=&\chi_0+\frac{\chi_0 \left(\lambda  \chi_0^2-2\mu_c^2 z_h^2-6\right)(z-z_h) }{4 z_h \left(\mu_I^2 z_h^2-2\right)}-
\nonumber\\
& &\frac{\chi_0(z-z_h)^2 }{16 z_h^2 \left(\mu_I^2 z_h^2-2\right)^2}\bigg\{z_h^2 \big [a_{2d} z_h \cos (\Pi_0)+2 \mu_I \sin\nonumber\\
& & (\Pi_0)\big]^2-\frac{3}{4} \chi_0^4 \lambda ^2+2 \chi_0^2 \lambda  \big\{z_h^2 \big[\mu_c^2+4\mu_I^2+\nonumber
\end{eqnarray}
\begin{eqnarray}
& &\mu_g^2 \big(2-\mu_I^2 z_h^2\big)\big ]+1\big\}+4 \mu_c^2 \mu_g^2 \mu_I^2 z_h^6-z_h^4 \big[\mu_c^4+ \nonumber\\
& &8 \mu_c^2 \mu_g^2+12 \mu_I^2 (\mu_c^2-\mu_g^2) \big ]-\nonumber\\
& &6z_h^2\left(\mu_c^2+4 \mu_g^2+8\mu_I^2\right)+15\bigg\}+\mathcal{O}[(z-z_h)^3],\nonumber\\
&& \\
\Pi(z) &=&\Pi_0+\frac{ (z-z_h)^2}{32\left(\mu_I^2 z_h^2-2\right)^2 }\bigg\{ \big [a_{2d}^2 z_h^2 -4\mu_I^2 \big] \sin (2 \Pi_0)-\nonumber\\
& &4 a_{2d} \mu_I z_h \cos (2 \Pi_0)\bigg \}+\mathcal{O}[(z-z_h)^3],\\
a_2(z) &=&a_{2d} (z-z_h)+\frac{(z-z_h)^2 }{4 z_h^2 \left(\mu_I^2 z_h^2-2\right)}\bigg \{a_{2d}\big [ g_5^2 \chi_0^2z_h \cos ^2\nonumber\\
& &(\Pi_0)+4  \mu_g^2 \mu_I^2 z_h^5-8  \mu_g^2 z_h^3+2  \mu_I^2 z_h^3-4  z_h\big ]+\nonumber\\
& &g_5^2 \mu_I \chi_0^2 \sin (2 \Pi_0)\bigg \}+\mathcal{O}[(z-z_h)^3],
\end{eqnarray}
\end{subequations}
where $\chi_0$, $\Pi_0$ and $a_{2d}$ are integral constants.
 In IR boundary expansions, the factor $1/(\mu_I^2z_h^2-2)$ exists in all the terms except for the leading one, which is related to the expression of temperature in Eq.~(\ref{temperature}). When temperature approach to zero, it will lead to the divergence of coefficients in IR boundary expansions. Therefore, in order to get reliable results at low temperature, more higher-order terms should be considered and the numerical steps should be properly chosen.

By using the UV and IR boundary expansion solutions, we can numerically solve the EOMs from both sides  with the ``shooting method''.
From the numerical solutions, we extract all the integral constants $\sigma$, $\pi_1$, $a_{2c}$, $\chi_0$, $\Pi_0$, and $a_{2d}$ from the solutions according to the holographic dictionary.
Through preliminary analysis, we find that the EOMs contain two independent solutions, corresponding to zero pion condensation $\pi_1=0$ ($\Pi(z)=0$) and finite pion condensation $\pi_1\neq 0$ ($\Pi(z)\neq 0$), respectively. Other than that, there is a  intermediate temperature region, in which it is relatively too high to form pion condensation but chiral condensation can arise. So that it is necessary to  separately discuss these two different solutions in the following contents.

 \subsection{Chiral condensation}\label{chiralcondensation}
  When pion condensation channel is turn off, in other words $\Pi(z)=0$, combining EOMs in Eqs.~(\ref{EOMb}) and (\ref{EOMd}), one has that $a_2(z)$ must be zero, too. The EOMs will degenerate to a simple  EOM of $\chi(z)$ as
\begin{eqnarray}\label{backgroundchi}
\chi ''+(3 A' -\Phi '-\frac{f'}{f})\chi '-\frac{e^{2 A}}{f}(-3  \chi-\mu_c^2 z^2  \chi+\frac{\lambda}{2}\chi^3)=0.\nonumber\\
\end{eqnarray}
The UV and IR asymptotic forms of the VEV $\chi(z)$ near $z=0$ and $z=z_h$ can be derived as
\begin{subequations}
\begin{eqnarray}
\chi(z)&=&m_q\zeta   z+\frac{\sigma  z^3}{\zeta }+\frac{1}{2} m_q  \zeta  z^3 \ln (z)\bigg(-\mu_c^2+2\mu_g^2+ \nonumber\\
& & \frac{\lambda}{2}  m_q^2 \zeta ^2\bigg)+\mathcal{O} (z^4),\label{backgroundchiboundary1}\\
\chi(z)&=&\chi_0+\frac{\chi_0 (z-z_h) \left(\chi_0^2 \lambda -2\mu_c^2 z_h^2-6\right)}{4 z_h \left(\mu_I^2 z_h^2-2\right)}+\mathcal{O}[(z-z_h)^3].\nonumber\\ \label{backgroundchiboundary2}
\end{eqnarray}
\end{subequations}

By using the ``shooting method'', we can numerically solve the Eq.~\eqref{backgroundchi}  with the boundary conditions given in Eqs.~\eqref{backgroundchiboundary1} and \eqref{backgroundchiboundary2} to study the crossover from $\chi$SB phase to $\chi$SR phase in terms of $T$ at fixed $\mu_{I, f}$. Then one can obtain the profiles of scalar VEV $\chi(z)$ from the numerical solutions.

\begin{center}
\tabcaption{\label{parameters} The parameters insure the self consistence of the meson spectra and the relevant decay constants in the IR-improved soft-wall AdS/QCD model. Case I accompanies a strange rising-up in the chiral condensate behavior. Case II is derived without consider the scalar meson spectrum. Taken from Ref.~\cite{Z.Fang}.}
\footnotesize
\begin{tabular*}{8.2cm}{c|c c c c c}
 \hline
Parameter &$m_q$(MeV)&$\mu_g$(MeV)&$\mu_c$(MeV) & $\lambda$&$g_5$\\
\hline
Case I &3.366 & 440 & 1180 & 33.6& $2\pi$\\
Case II &3.22 & 440 & 1450 & 80& $2\pi$\\
\bottomrule
\end{tabular*}
\end{center}

For the IR-improved soft-wall AdS/QCD model proposed in Ref.~\cite{Z.Fang}, there are two different sets of parameters, including $m_q$,  $\mu_g$, $\mu_c$, $\lambda$ and $g_5$, Case I and Case II, as shown in Tab.~\ref{parameters}. By using the parameters of Case I, the meson spectrum can be well reproduced. As to case II, the meson spectrum expect for the scalar meson spectrum can be well matched, other than that the $\pi-\rho$ coupling constant and the decay constants of $\pi$, $\rho$, and $a_1$ are more consistent with experimental data.

For these two cases, we separately study the behaviors of chiral condensation $\sigma(\mu_{I,f},T)$, as shown in Figs.~\ref{figchi}(a) and~(b). In Fig.~\ref{figchi}(a), the curves of $\sigma(\mu_{I, f},T)$  almost keep saturate value in the low temperature region along with small unphysical bumps, and then smoothly decrease while keeping $T$ increasing.  Notice that the small unphysical bump behaviors also can be found in other holographic models~\cite{P.Colangelo}.
 Because the small bump also exist in small $\mu_I$ region, the curves of $\sigma(\mu_{I, f},T)$, with a set of different $\mu_{I,f}$, cross each other in low temperature region. The pseudo critical temperature of the chiral crossover transition is identified by the position of the peak of the susceptibility, $d^2\sigma/{d T}^2=0$. The measurement results are $T_c = 0.079$, $0.100$, and $0.117$ GeV for $\mu_{I,f}= 0.400$, $0.250$, and $0.050$ GeV, respectively. There is a tendency that the higher the pseudo critical temperature, the smaller $\mu_{I,f}$, which obey the mechanism that  $\mu_I$ and $T$  are both in favor of restoring the chiral symmetry.

\begin{center}
\centering
\includegraphics[scale=0.58]{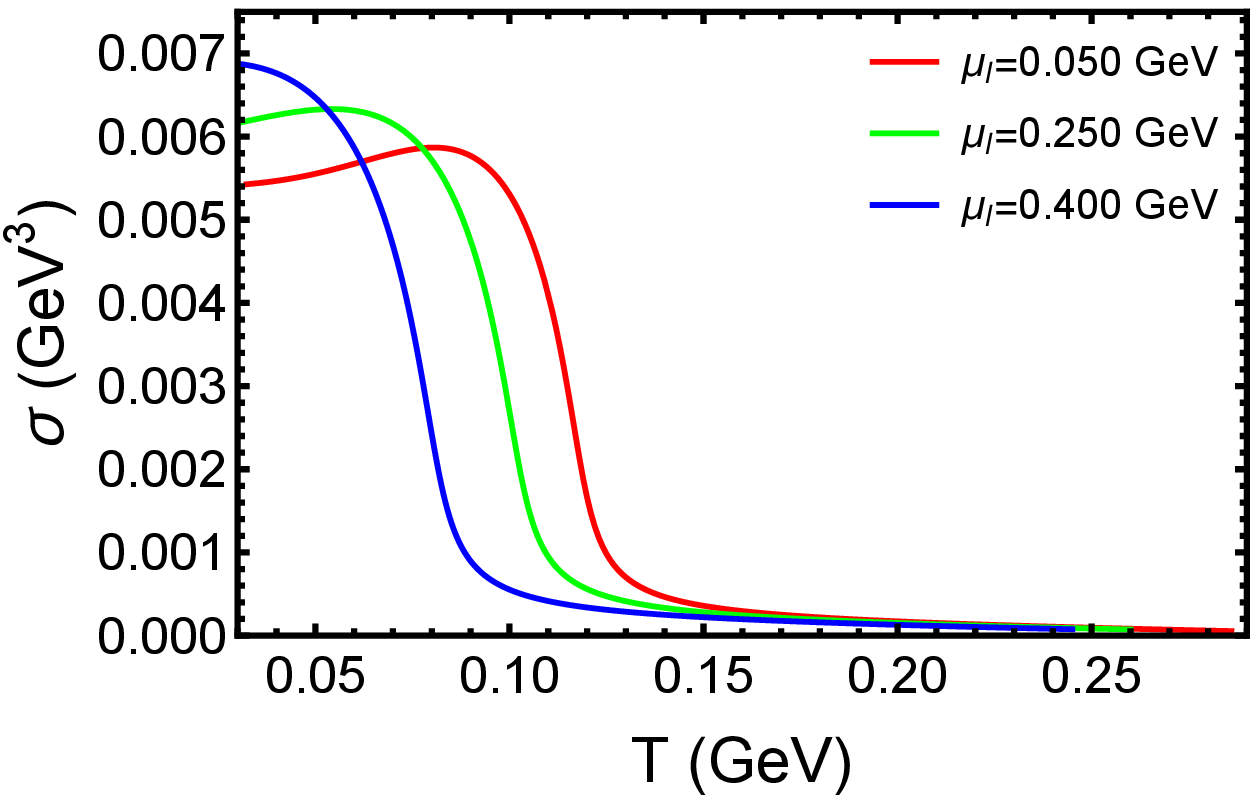} 
 \put(-88,33){\begin{overpic}[scale=0.23]{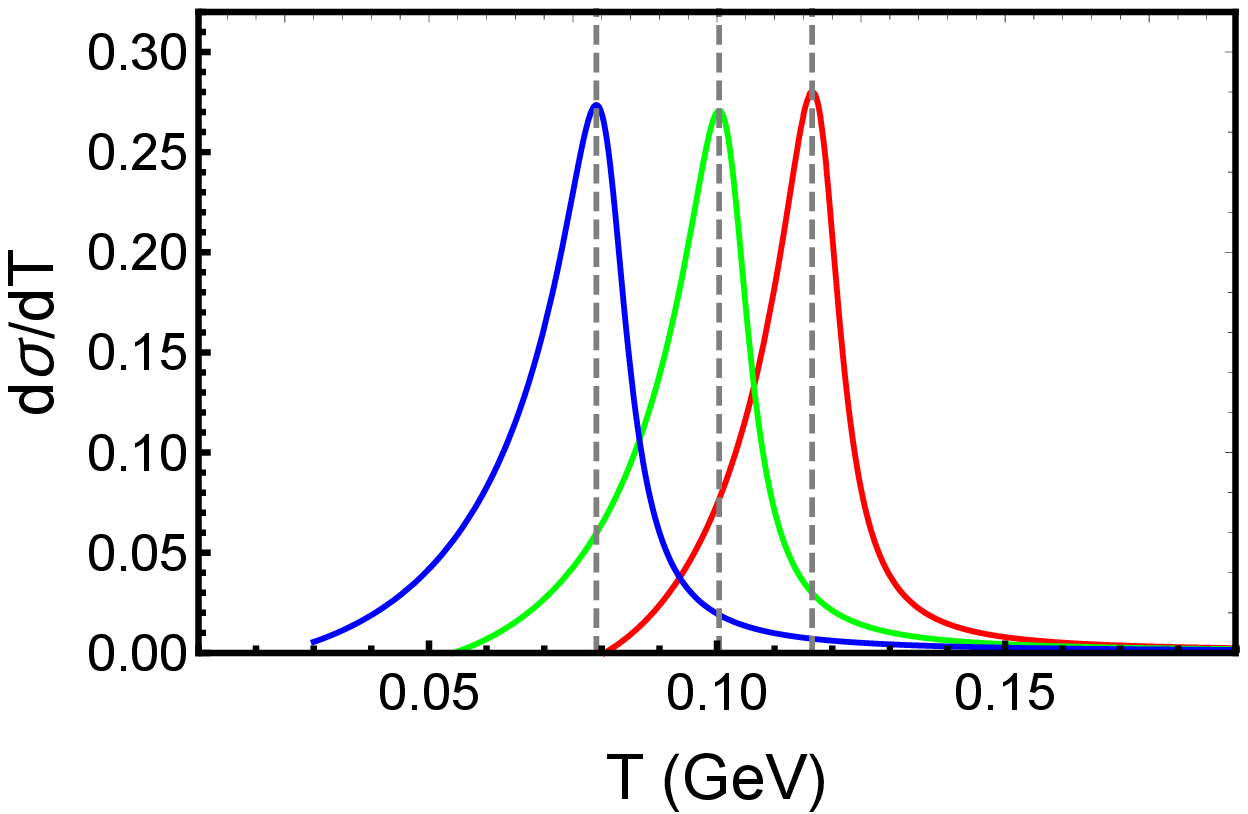} \end{overpic}}
 \put(-158,33){\large \color{black}{\bf (a)}}
\hfill
 \includegraphics[scale=0.58]{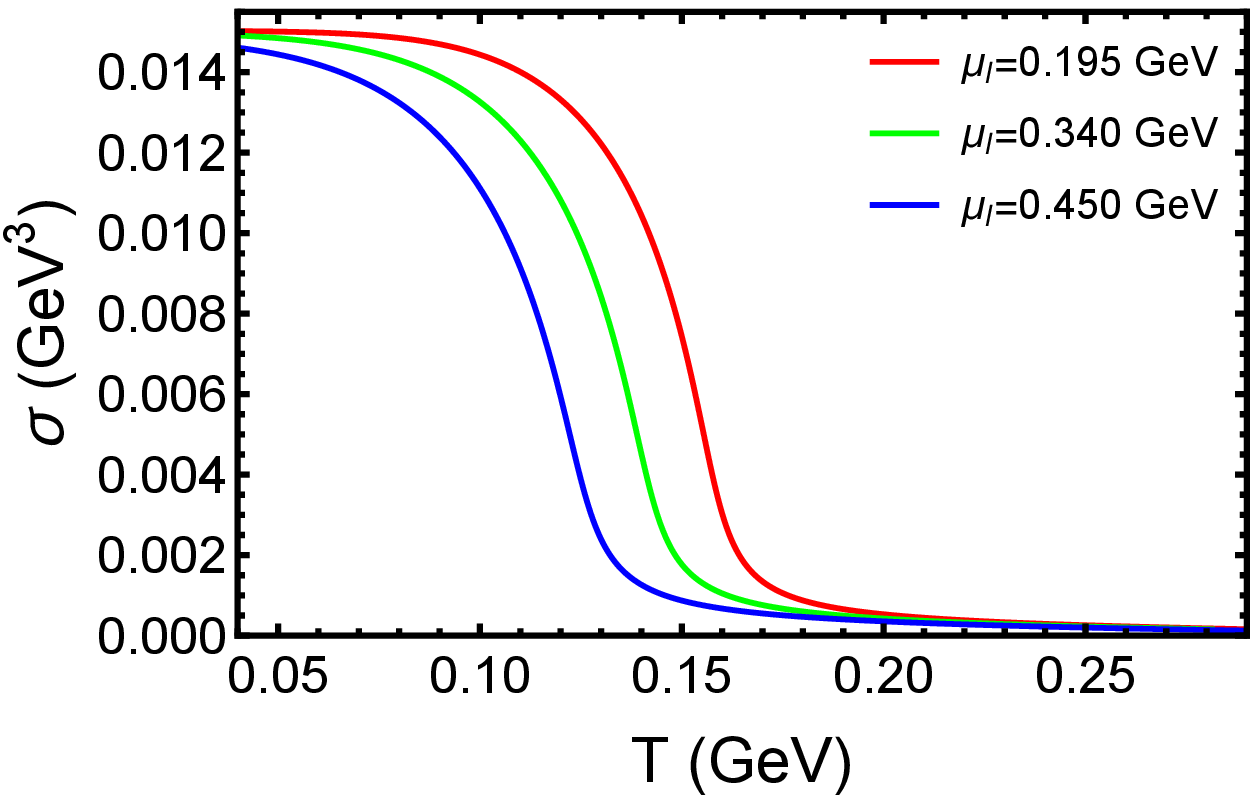} 
 \put(-88,33){\begin{overpic}[scale=0.23]{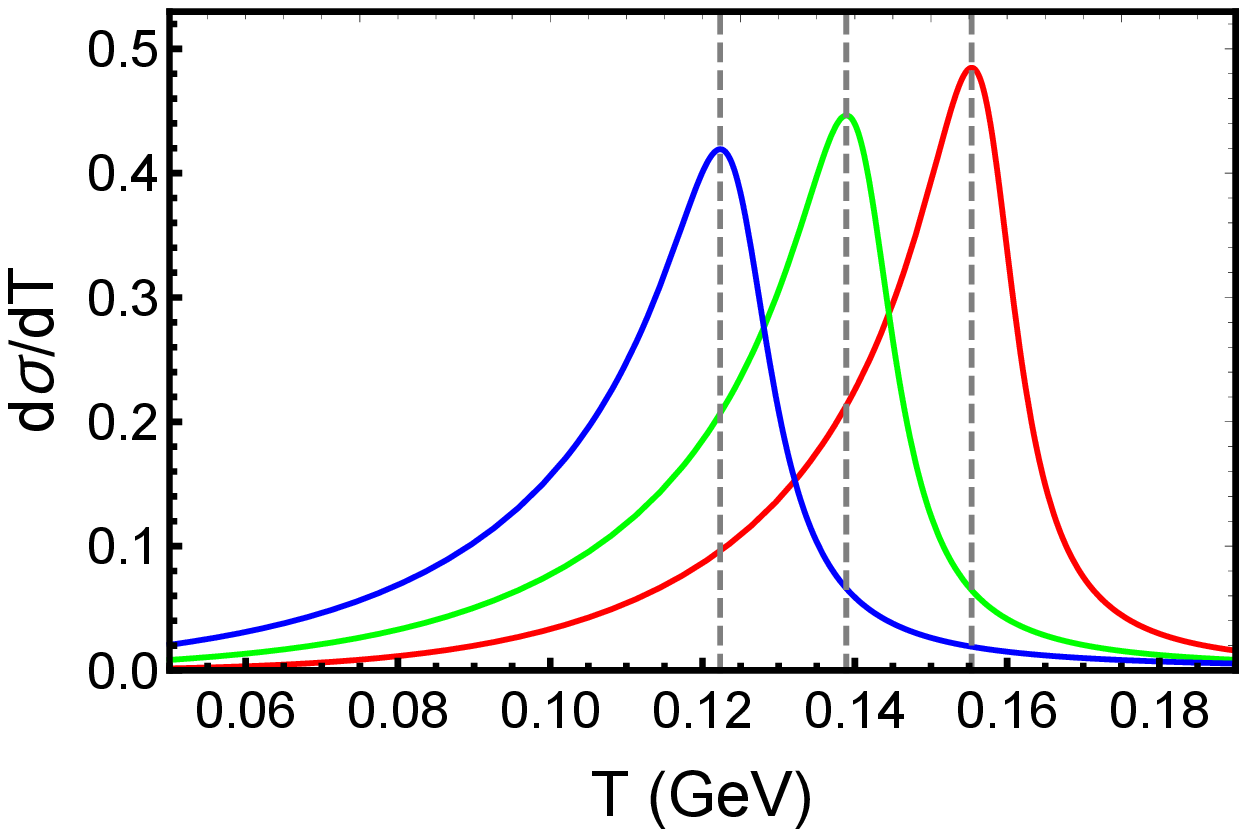} \end{overpic}}
  \put(-158,33){\large \color{black}{\bf (b)}}
\hfill
\figcaption{\label{figchi}(a) and (b), the dependence of chiral  condensation $\sigma(\mu_{I,f}, T)$ and susceptibility $d\sigma/dT$ (the inside figure) on temperature with the parameters in Case I and Case II, with three different fixed isospin chemical potential $\mu_{I,f}$, respectively. Where the pseudo critical point is determined by the position of the peak of susceptibility, $d^2\sigma/{d T}^2=0$}
\end{center}

 In Fig.~\ref{figchi}(b), the behaviors of $\sigma({\mu_{I, f}, T})$ is generally consistent with with case I, except for the strange rising-up behavior in low temperature region. The pseudo critical temperatures are $T_c=0.122$, $0.139$, and $0.155$ GeV for $\mu_{I, f}=0.195$, $0.340$, and $0.450$ GeV, respectively. However, when  $\mu_{I, f}$ is lower, the peak susceptibility is larger, but there is no such significant difference in Case I. In addition, when $\mu_B$ and $\mu_I$ are both zero, the pseudo critical temperatures are $T_{c,1}= 0.117$ GeV in Case I and $T_{c,2}=0.164$ GeV in Case II, in which the latter is consistent with results from LQCD simulations~\cite{PhysRevD.85.054503, Bors2010}.   All these behaviors indicate the parameters of Case II is more self-consistently. Therefore, we choose parameters in Case II for our following studies in this paper.

 \subsection{Pion condensation}\label{pioncondensation}
When $\mu_I$ is large enough,
 the $U_I(1)$ symmetry is spontaneously breaking with a massless Goldstone boson~\cite{P.F.Zhuang}. Although symmetry analysis can give a profile of the phase transition, the detailed properties with finite $\mu_I$ and $T$ are still ambiguous. In this subsection, we will study pion condensation and chiral condensation, as well as their interdependent behaviors.

\begin{center}
\includegraphics[scale=0.58]{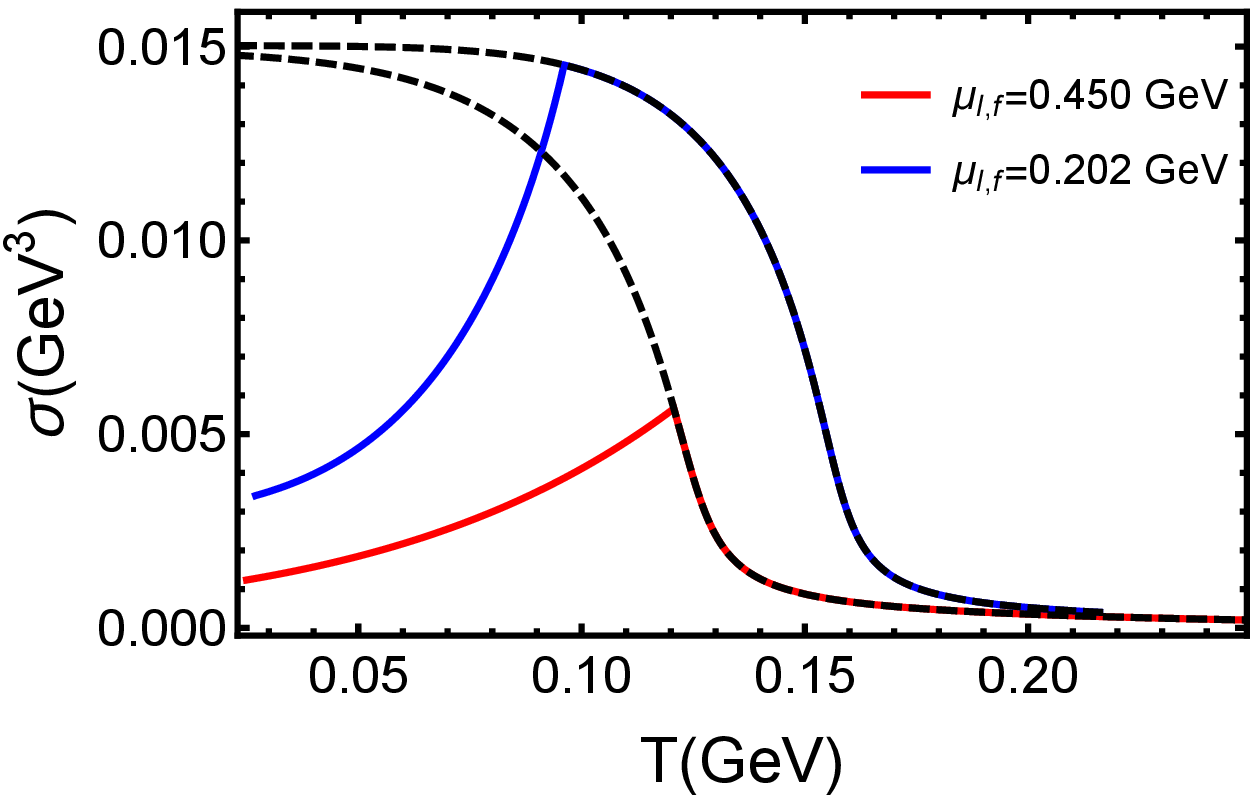} 
 \put(-23,33){\large \color{black}{\bf (a)}}
 \hfill
\includegraphics[scale=0.58]{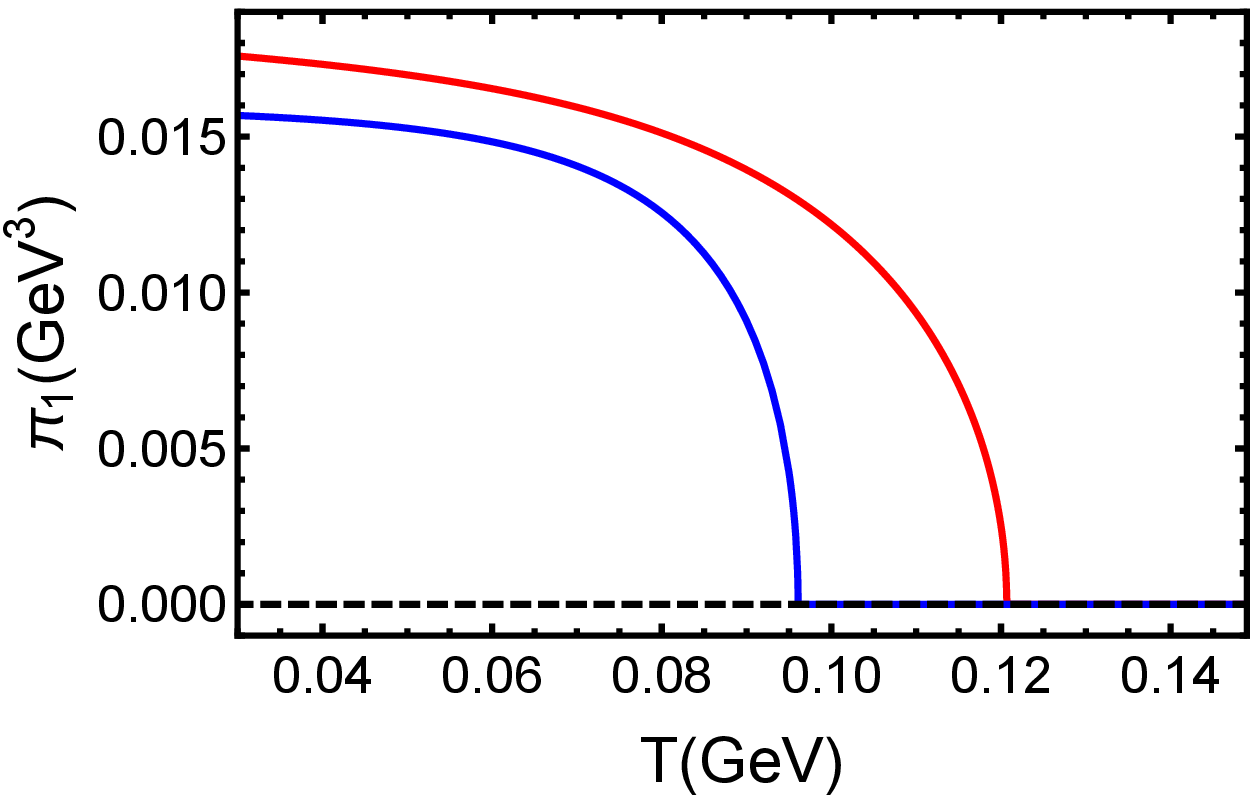}
 \put(-23,35){\large \color{black}{\bf (b)}}
 \hfill
 \includegraphics[scale=0.8]{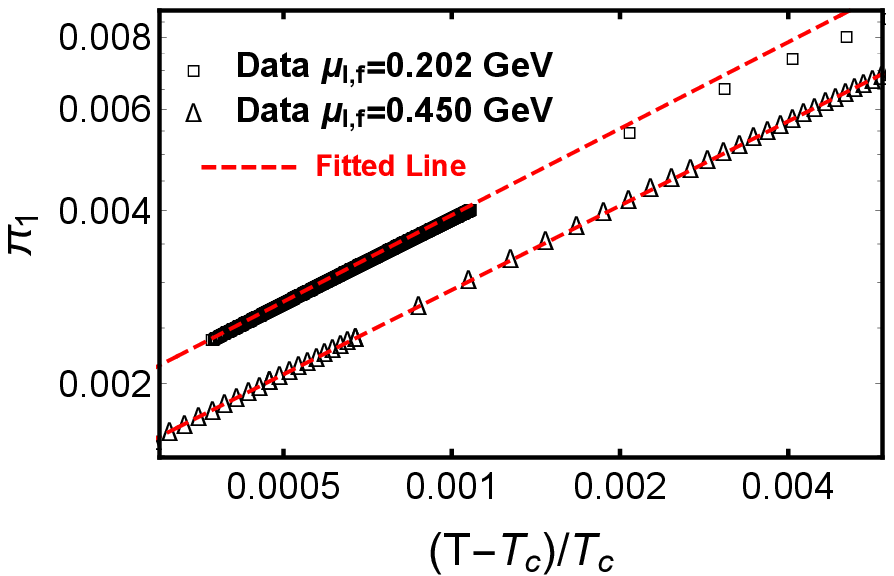} 
 \put(-21,33){\large \color{black}{\bf (c)}}
\figcaption{\label{figchipit}  Chiral condensation (a) and pion condensation (b), with fixed isospin chemical potential $\mu_{I,f}$. The dashed line and solid line correspond to $\pi_1=0$ and $\pi_1\neq 0$, respectively. The critical temperature of pion condensation are $T_{c,\pi_1}=0.0961$  and $0.121$ GeV for  $\mu_{I,f}= 0.202$, and $0.450$ GeV, respectively. (c) Red lines fit the data in the critical region with $\pi_ 1 =
 504.704 (0.096 - T)^{0.499}$, and $\pi_ 1 = 343.333 (0.121 - T)^{0.487} $ for $\mu_{I,f}= 0.202$, and $0.450$ GeV, respectively.}
\end{center}

  In Fig.~\ref{figchipit}, we  numerically solve the complete EOMs of Eqs. ~\eqref{EOMa}, \eqref{EOMb} and \eqref{EOMd} to study the properties of pion condensation $\pi_1(\mu_{I, f}, T)$ and chiral condensation $\sigma(\mu_{I, f}, T)$ in terms of $T$ with fixed $\mu_{I,f}$.  The behaviors of $\pi_1(\mu_{I, f}, T)$ are shown in Fig.~\ref{figchipit}(b), along with the increasing of  $T$,  $\pi_1(\mu_{I, f}, T)$ keeps continuously decreasing all the way down to zero at critical points $T_{c, \pi_1}$, in which $T_{c,\pi_1}=0.096$ and $0.121$ GeV corresponding to $\mu_{I,f}=0.202$ and $0.450$ GeV, respectively. It seems that it is a second order phase transition, in order to verify the universality classes of the pion condensation, we numerically fit out the critical exponent  and the results are $\beta=0.499$ and $0.487$, corresponding to $\mu_{I,f}=0.202$ and $0.450$, respectively, as shown in Fig.~\ref{figchipit}(c). These results are very close to $1/2$, which means the pion condensation in this model belongs to the class of 4D mean field.  There probably have two reasons for it, on one hand, the exact holographic duality is based on the assumption of large $N_c$ limit, however $N_c=3$ is just a rough approximation; on the other hand, the back reaction of condensations to the background is also ignored in this AdS/QCD model, and teat the solution of AdS/RN black hole as the bulk background.  The curves of $\sigma(\mu_{I,f}, T)$ are shown in Fig.~\ref{figchipit}(a), in which solid lines and dashed lines represent for the solutions with or without pion condensation, respectively. Comparing to $\pi_1(\mu_{I,f},T)$ in Fig.~\ref{figchipit}(b), we can divide the $\mu_I$-axis into two regions by $T_{c,\pi_1}$. When $\mu_I\geq\mu_{I, \pi_1}^c$, these two solutions collapse into one; when $0<\mu_I<\mu_{I, \pi_1}^c$, these two solutions, respectively, rises up and drops down with the pion condensation channel turn on or off, which means the chiral condensation is depressed by the pion condensation.


\begin{center}
 \includegraphics[scale=0.58]{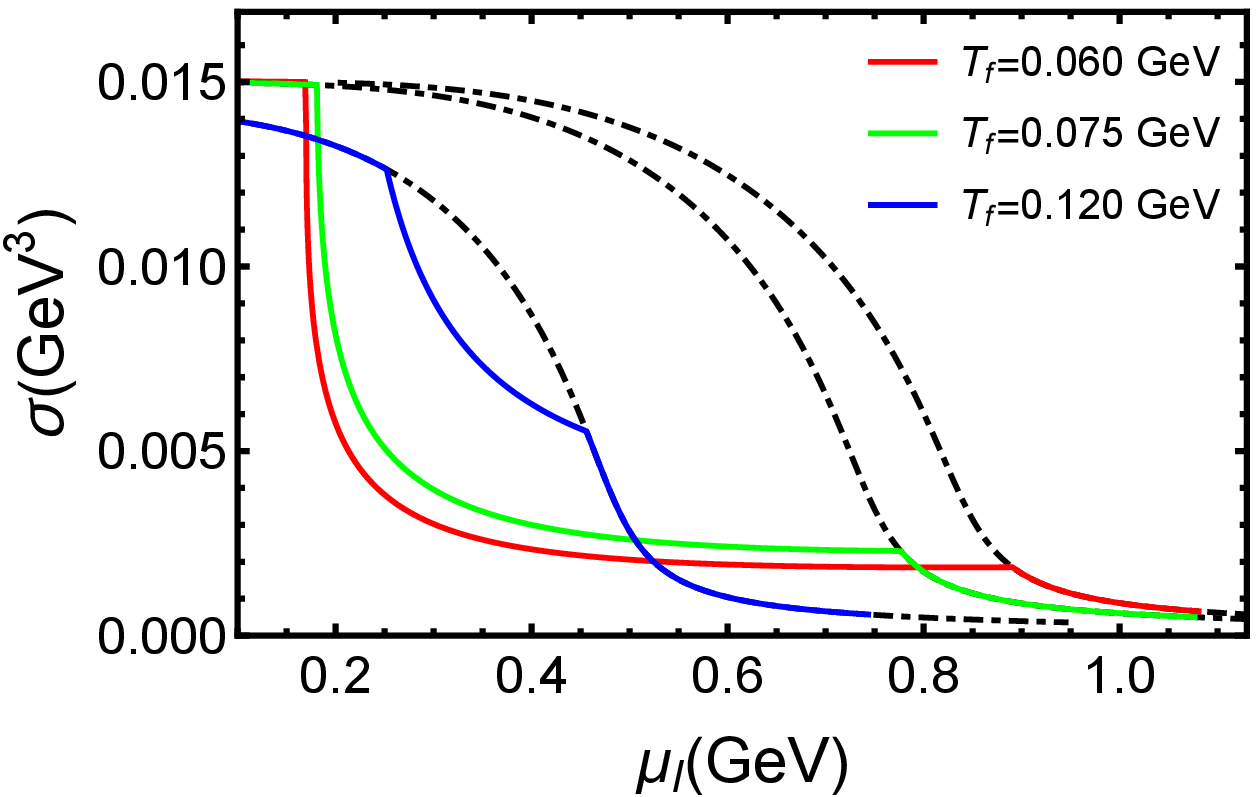} 
 \put(-21,35){\large \color{black}{\bf (a)}}
\hfill
\includegraphics[scale=0.58]{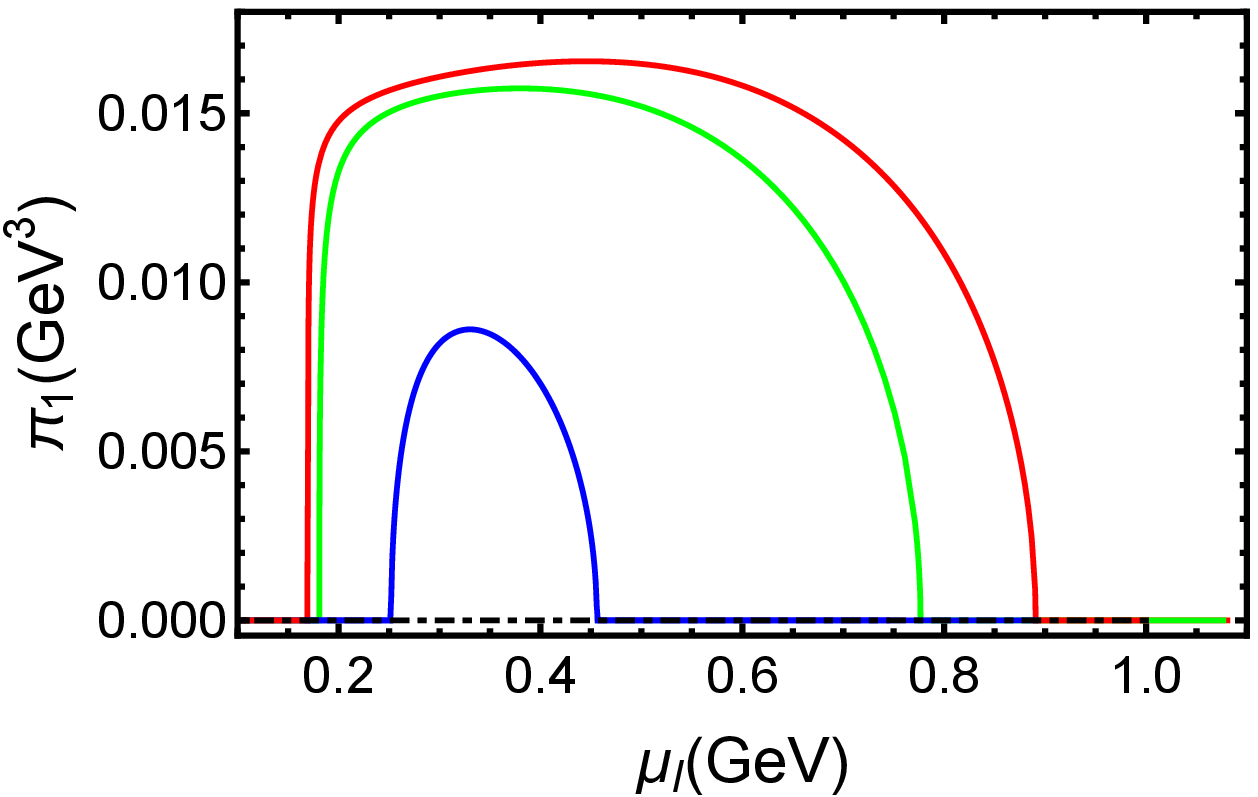}
 \put(-22,35){\large \color{black}{\bf (b)}}
\figcaption{\label{figchipi}  Chiral condensation (a) and pion condensation (b), with fixed temperature $T_f$. The dashed line and solid line correspond to $\pi_1=0$ and $\pi_1\neq 0$, respectively. The left and right critical point of pion condensation are $\mu_{I,L}^c=0.170,\ 0.181$, and $0.251$ GeV, and $\mu_{I,R}^c=0.457,\ 0.777$, and $0.891$ GeV, corresponding to $T_f=0.060,\ 0.075$, and $0.120$ GeV, respectively.}
\end{center}

\begin{center}
\includegraphics[scale=0.5]{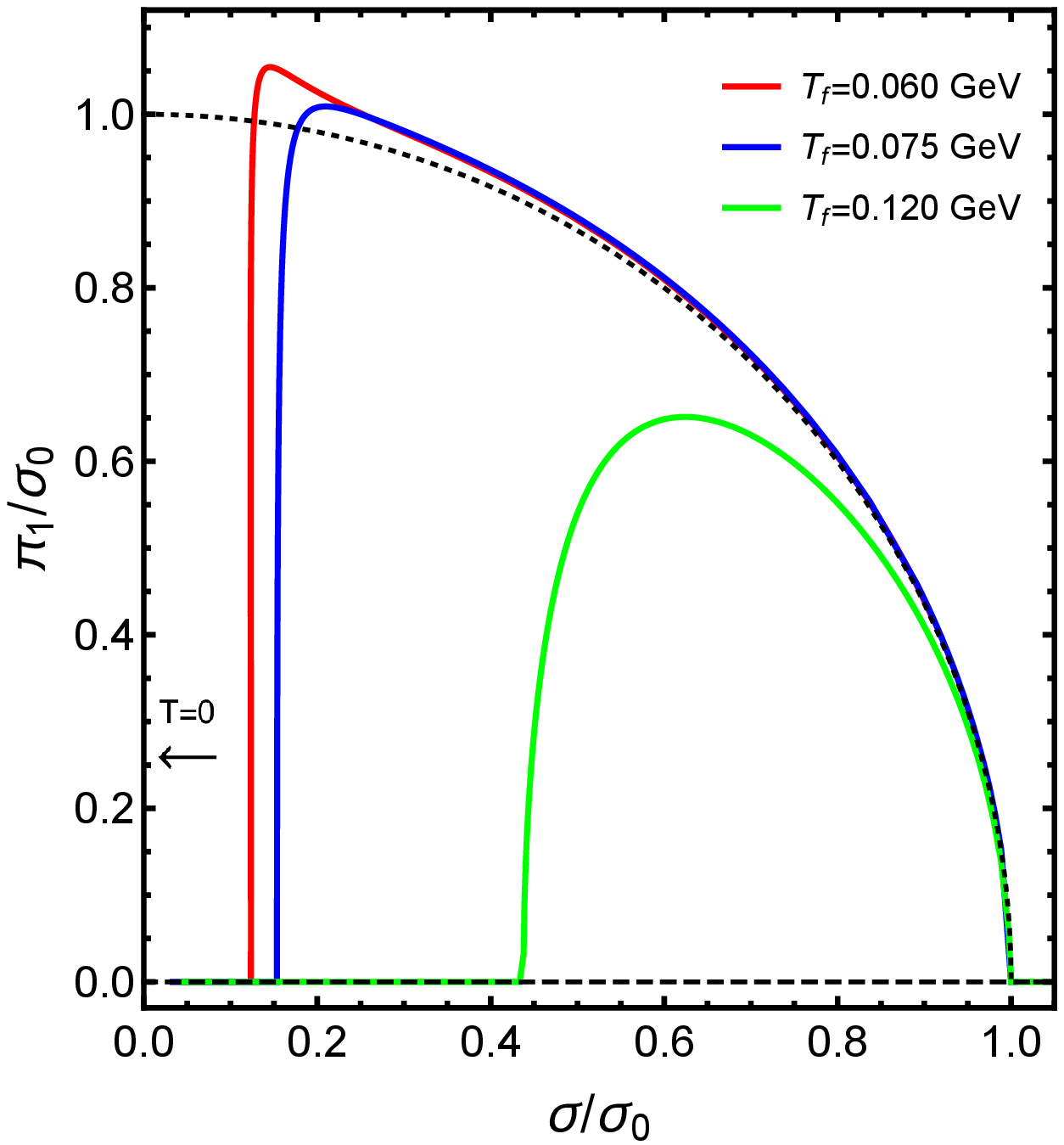} 
\figcaption{\label{figenhancement} The interdependence relationship  between ${\sigma}/{\sigma_0}$ and $\pi_1/\sigma_0$, in which $\sigma_0=\sigma(\mu_{I,L}^c,T_f)$. The dashed black curve represents for the unit circle $ \widetilde{\sigma}/\sigma_0=1$. The sharp decreasing line infinitely trends to y-axial as $T\rightarrow 0$.}
\end{center}
However, pion condensation can also be studied through another perspective, investigating the dependence of $\pi_1(\mu_I, T_f)$ on $\mu_I$ with different fixed temperatures $T_{f}$. The numerical results are shown in Fig.~\ref{figchipi}, in which solid lines and black dashed lines represent  turning on and off the pion condensation channel, respectively. In Fig.~\ref{figchipi}(b), pion condensation possesses two critical points ($\mu_{I,L}^c$ and $\mu_{I,R}^c$), which divide the figure into three areas, one non-zero pion condensate region ($\mu_{I,L}^c<\mu_I<\mu_{I,R}^c$) in the middle and two zero pion condensate regions ($0<\mu_I\le\mu_{I,L}^c$ and $\mu_I\ge\mu_{I,R}^c$) on both sides. The critical points are $\mu_{I, L}^c=0.170$, $0.181$ and $0.251$ GeV on the left and  $\mu_{I,R}^c=0.457$, $0.777$, and $0.891$ GeV on the right for $T_{f}=0.060$, $0.075$ and $0.120$ GeV, respectively. We find that along with the increasing of $T_f$, $\mu_{I, L}^c$ and $\mu_{I, R}^c$ close to each other and the pion condensate is gradually decreasing, which indicate that there exists a critical temperature at which pion condensate just disappears in all $\mu_I$ region and the pion condensation phase should possess a raised area in the space of $\mu_I$ and $T$.
In the corresponding middle region (non-zero pion condensation region) of Fig.~\ref{figchipi}(a), the chiral condensate is depressed, and the degrees of depression is relatively proportional to the strength of pion condensation. In the regions on both sides,  $\sigma_1(\mu_T, T_f)$ behaves the same as the ordinary chiral crossover without pion condensation.
Finally, we note that the behavior that $\pi_1(\mu_{I}, T_f)$ with both left and right critical points also shows  in Ref.~\cite{J.B.Kogut}, calculated by LQCD, and  Ref.~\cite{D.Li} by a soft-wall AdS/QCD model.

The chiral condensation and pion condensation have separately studied in the preceding part of this section. However, there is mutual-interaction between themselves. Therefore it is necessary to investigate their interdependence relationships.
Figure~\ref{figenhancement} shows the dependency between $\sigma(\mu_I,T_f)/\sigma_0$ and \\$\pi_1(\mu_I,T_f)/\sigma_0$ , in which $\sigma_0=\sigma(\mu_{I,L}^c,T_f)$, the dashed black curve represents for the unit circle $ \widetilde{\sigma}/\sigma_0=1$, and the absolute chiral condensation is defined as $\widetilde{\sigma}=\sqrt{\sigma^2+\pi_1^2}$. Along the increasing of $\mu_I$, the green curve ($T_f=0.120$ GeV) falls into the circle, the red and blue curves ($T_f=0.060$ and $0.075$ GeV) first show the enhancement behavior and then drop sharply to zero when $\mu_I$ is large enough.  These curves indicate a tendency that when temperature is infinitely close to zero, the dropping line will be infinitely close to the $\pi_1/\sigma_0$-axes and the enhancement tends to increase to infinitely as $\mu_I$ increases, and the enhancement only arise when the temperature is low enough. From the expression of action and dilaton, we know that when temperature tends to zero, the dilaton term will approach to one as $z_h$ approach to infinity, the soft wall boundary will back to the hard wall cut, therefore the zero temperature asymptotic behaviors qualitatively coincide with the hard wall results in Ref.~\cite{enhancement1}.

\section{Phase diagram}\label{phasediagram}
To obtain the complete phase diagram on the $\mu_I$ and $T$ plane, we can let the fixed $T_f$ ergodic the entire $T$- axis, and case by case solve the EOMs in Eqs.~\eqref{EOM} at a fixed $T_f$. Then we can extract the critical point of pion condensation, as well as the pseudo critical point of chiral condensation from the solutions, as we did in Secs.~{\ref{chiralcondensation}} and {\ref{pioncondensation}}.


However, if we just want to determine the phase boundaries, case by case solving the full EOMs may be not necessary. Here we will introduce a more direct method.  From the EOMs in Eq.~(\ref{EOM}) and the UV expansions solution in Eq.~(\ref{UVexpansion}), we know that $\Pi(z)$ and $a_{2}(z)$ are vanishing small when $\pi_1$ and $a_{2c}$ approaching to zero. In addition, through the researches in Sec.~{\ref{pioncondensation}}, we know that $\pi_1$ and $a_{2c}$\footnote{The numerical results of $a_{2c}$ are gotten simultaneously with $\pi_1$ and $\sigma$, but we only concern about the properties of $\pi_1$ and $\sigma$ and we just dismiss it in the main text.} changes continuously from zero to non-zero values around the critical point ($\mu_{I}^c$, $T_c$), so that we can expand $\Pi(z)$ and $a_{2}(z)$ in the critical region, based on the background of  $\chi(z)$.  Thus, when there is an infinitesimal perturbation near the critical point$(\delta \mu_I, \delta T)= (\mu_I-\mu_I^c, T-T_c)$, we expand $\Pi(z)$ and $a_2(z)$ and just keep to linear terms, which is good enough to satisfy the EOMs, and then the boundary EOMs are derived as
\begin{subequations}\label{phaseboundary}
\begin{eqnarray}
& &\delta\Pi ''+\delta\Pi ' \left(3 A'+\frac{f'}{f}-\Phi '+\frac{2 \chi '}{\chi }\right)-\frac{v (\delta a_2-\delta\Pi  v)}{f^2}=0,\nonumber\\
&& 
\end{eqnarray}
\begin{eqnarray}
& &\delta a_2''+\delta a_2' \left(A'-\Phi '\right)-\frac{g_5^2 e^{2 A} \chi ^2 (\delta a_2-\delta\Pi  v)}{f}=0.\nonumber\\
\end{eqnarray}
\end{subequations}
From Eq.~\eqref{phaseboundary}, we get IR-boundary conditions as
\begin{subequations}\label{IRboundary1}
\begin{eqnarray}
\delta\Pi(z)&=&\Pi_0+\frac{\mu_I  (z-z_h)^2 (a_{21} z_h-2 \mu_I  \Pi_0)}{8 \left( \mu_I ^2 z_h^2-2\right)^2}z^2+\nonumber\\
& &\mathcal{O}[(z-z_h)^3],\\
\delta a_2(z)&=&a_{21} (z_h-z)-\frac{(z-z_h)^2 }{4 z_h^2 \left(\mu ^2 z_h^2-2\right)}\big\{a_{21} z_h \big[\chi_0^2 g_5^2+\nonumber\\
& &2 \left(\mu_I ^2 z_h^2-2\right) \left(2 \mu_g^2 z_h^2+1\right)\big ]-2 \chi_0^2 g_5^2 \mu_I  \Pi_0\big\}+\nonumber\\
& &\mathcal{O}[(z-z_h)^3],
\end{eqnarray}
\end{subequations}
where $\Pi_0$ and $a_{21}$ are integral constants.

\begin{center}
\includegraphics[scale=0.6]{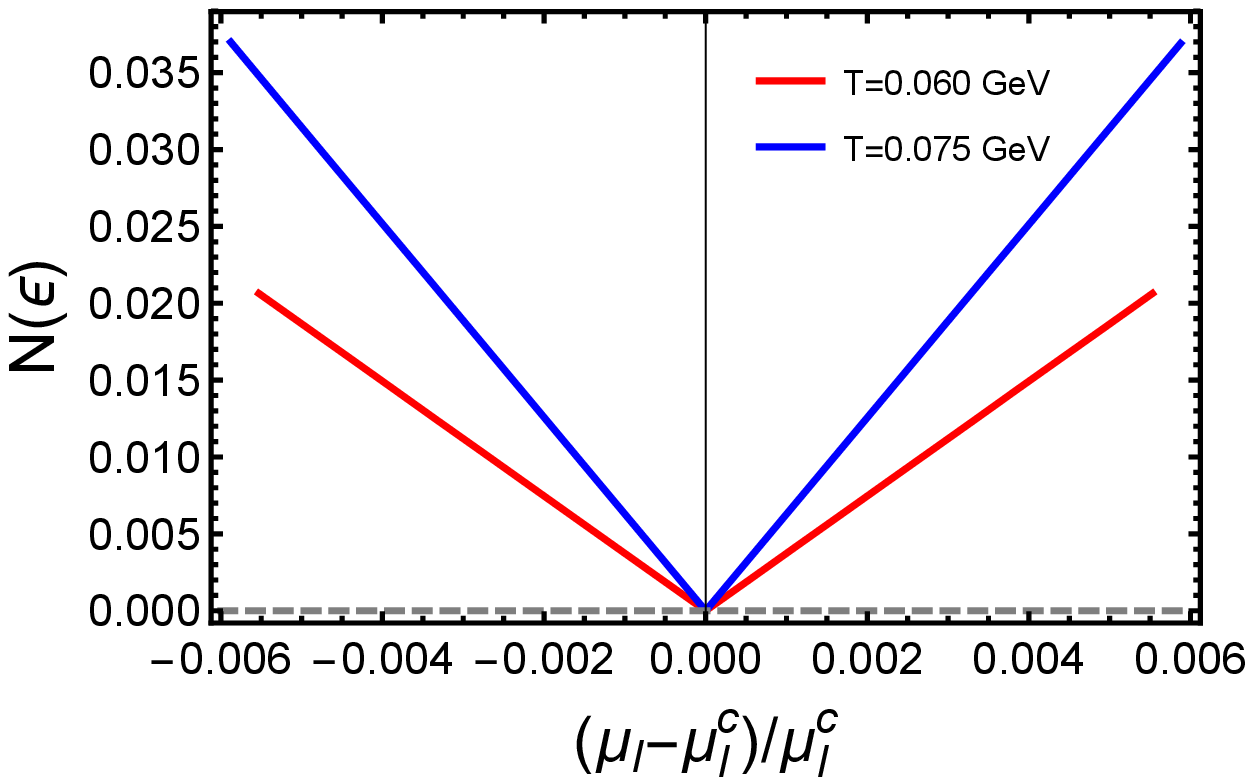} 
\figcaption{\label{linear} The Norm $N(\epsilon)= \sqrt{\delta\Pi[\epsilon]+\delta a_2[\epsilon] }$ is linearly proportional to $\delta \mu_I=\left |(\mu_I-\mu_I^c)/\mu_I^c\right |$, in which critical points $\mu_{I,L}^c=0.170$ and $0.181$ GeV correspond to $T_{f}=0.060$ and $0.075$ GeV, respectively.}
\end{center}

The boundary EOMs in Eq.~\eqref{phaseboundary}, is a set of
linear second order differential equations, so that we can set $\Pi_0=1$, and numerically solve the boundary EOMs by using the IR-boundary conditions in Eq.~\eqref{IRboundary1}. Only when $(\mu_I,T)=(\mu_I^c,T_c)$, the conditions of $\delta \Pi(z)|_{z=0}=0$ and $\delta a_2(z)|_{z=0}=0$ can be simultaneously satisfied. In Fig.~\ref{linear}, we numerically test the dependence of the norm  $N(\epsilon)=\sqrt{\delta \Pi(\epsilon)^2+\delta a_2(\epsilon) ^2}$ on $\delta\mu_I$, in which $\epsilon$ is an infinitely small number, and it behaves linearly proportional to $\delta \mu_I=\left |(\mu_I-\mu_{I, L}^c)/\mu_{I, L}^c\right |$, where $\mu_{I, L}^c\approx 0.170$ and $0.181$ GeV corresponding to $T_f=0.060$ and $0.075$ GeV, respectively.

On the other hand, the $U_I(1)$ symmetry is spontaneously broken by the pion condensation, which generates a massless Nambu-Goldstone boson. To identify the pion mass on the phase boundary, we can analyze the goldstone mode in the momentum space, $q=(\omega, \bold q)$. We expand the Lagrange in Eq.~\eqref{effectlagrangian} to squared terms to deduce the EOMs in the momentum space as
\begin{subequations}\label{pEOMs}
\begin{eqnarray}
& &a_2' \left(A'-\Phi '\right)+a_2''-\frac{ {g5}^2 e^{2 A} \chi ^2 \left(a_2+\Pi  (\omega -v)\right)}{f}=0,\nonumber\\
\\
&&\Pi ''+\Pi ' \left(3 A'+\frac{f'}{f}-\Phi '+\frac{2 \chi '}{\chi }\right)+\nonumber\\
&&\frac{(\omega -v) \left [a_2+\Pi  (\omega -v)\right ]}{f^2}=0,
\end{eqnarray}
\end{subequations}
in which we have used the condition that three dimensional momentum $\bold q=0$. The massless goldstone mode is the eigenstate of EOMs in the momentum space ~\cite{colangelo2012medium,PhysRevD.81.065024,PhysRevD.80.035001} with $\omega=m_{\pi_1}=0$ and $\mu_I=\mu_I^c$.
In other words, the pion becomes massless and pion condensate forms when $\mu_I$ increase to $\mu_I^c$.  Under this condition,  Eqs.~\eqref{pEOMs} indeed coincide with the boundary EOMs in Eqs.~\eqref{phaseboundary}. It proves that on the pion condensation phase boundary the mass of the pion equals to zero, $m_{\pi_1}(T_c, \mu_I^c)=0$.

\begin{center}
\includegraphics[width=7.8cm]{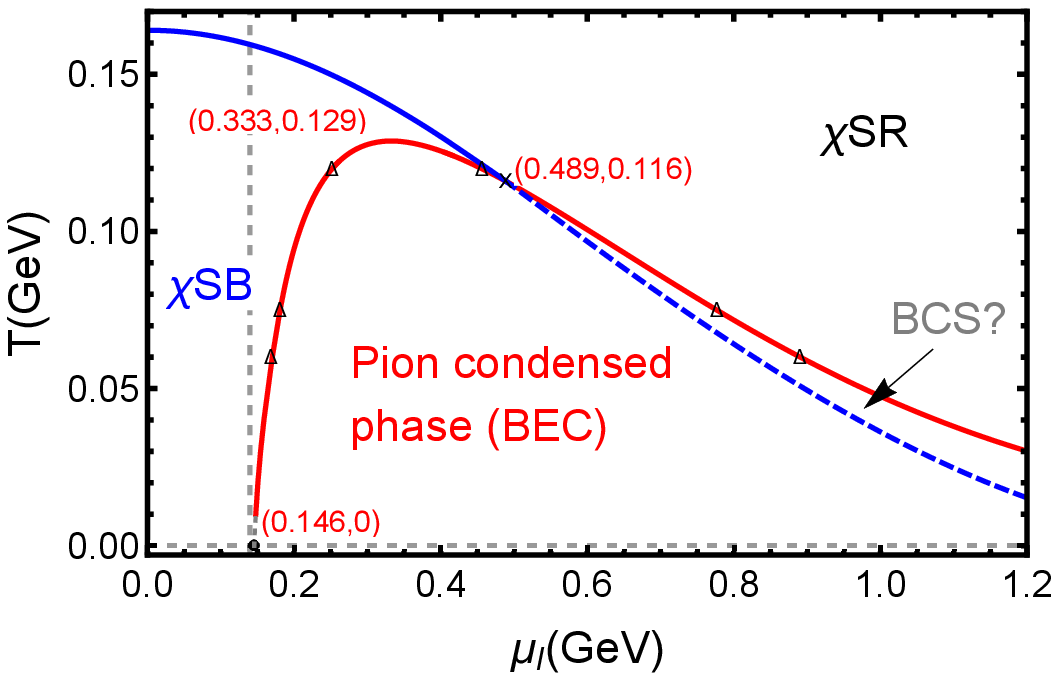}
\figcaption{\label{figphasediagram} The phase transition diagram of chiral crossover and pion condensation in $\mu_I-T$ plane. The blue curve is the boundary between normal $\chi$SB phase and $\chi$SR phase. The Red curve and the $\mu_I$-axis surround the pion condensed phase. Gray dashed curve is the extrapolated pion condensation boundary in very low temperature region and the end point is the critical point, $\mu_{I,T=0}^{c}=0.146$ GeV, at zero temperature.
The chiral crossover boundary and the pion condensation boundary meet at the tricritical point $(\mu_{I,tri}^c, T_{c,tri})=(0.489,0.116)$ GeV. The extreme point of pion condensation boundary is $(\mu_{I,top}^c, T_{c,top})=(0.333,0.129)$ GeV.
 We suppose the blue curve below the tricritical point indicates a BEC-BCS crossover. The black triangle points are the critical points of pion condensation, which are  obtained by the method in Sec.~{\ref{pioncondensation}}.}
\end{center}

By solving the boundary EOMs, we obtain the compelet phase diagram on the $\rm \mu_I$-$\rm T$ plane, as shown in Fig.~\ref{figphasediagram}.  The pion condensed phase region is surrounded by the red line and x-axis and is a convex shape with a top point $(\mu_{I,top}^c,T_{c,top})=(0.333, 0.129)$ GeV. The black triangle points on the red line are obtained by solving the complete EOMs, and this consistency proves that the two methods give the same results. Despite the divergence problem of the expansion coefficients at zero temperature, we can still infer zero temperature critical point $\mu_{I, T=0}^c$  from the  low temperature phase boundary trend, and we have $\mu_{I, T=0}^c\approx 0.146$ GeV,  which is represented by the end point of the gray dashed line.  The zero temperature critical point coincides well with the results of $\mu_{I, T=0}^c=m_0$ in the numerical LQCD~\cite{B.Brandt}, analytical chiral perturbation method~\cite{D.T.Son1, P.Adhikari}, as well as hard wall AdS/QCD~\cite{enhancement1}.  In large $\mu_I$ region, the pion condensation boundary has a tendency of infinitely approaching to zero temperature and infinite $\mu_I$, which means  pion condensation loses its right side critical point at zero temperature and it is mutual verified with the tendency of enhancement studied  in Sec.~{\ref{pioncondensation}}. The blue line respect for the boundary of  chiral crossover, and it comes across pion condensation boundary at a tricritical point, at which three phases (pion condensation phase, normal $\chi$SB phase and $\chi$SR phase) coexistence terminates, $ (\mu_{I,tri}^c,T_{c,tri} )$ $= (0.489, 0.116)$ GeV.  The part of chiral crossover boundary, which is in the pion condensation region, apart the pion condensation region into left and right areas. The chiral crossover accompany with the chiral symmetry restoration can be seen as a signal of color deconfinement transition(if one supposes that the two transition is coincident). So it indicates that left part is a pion condensed phase (BEC), however the right part, to a certain degree, is a color deconfinement  Bardeen-Cooper-Schrieffer (BCS) phase, which is a Fermi liquid with cooper pairing is formed as a consequence of an attractive interaction between quarks in the isospin channel.

\section{Conclusion and discussion}\label{conclusion}
In this paper, we have studied pion condensation and chiral condensation with finite $\mu_I$ and finite $T$ in the IR improved soft-wall AdS/QCD model. Under a fixed $T_f$, we find that pion condensation exists two  critical points separately located in small $\mu_I$ region and large $\mu_I$ region, similar behaviors also show in the LQCD in Ref~\cite{J.B.Kogut}, the NJL model in Ref~\cite{P.F.Zhuang} and a solf-wall model in Ref~\cite{D.Li}.
The behaviors of the pion condensation continuous changes from zero to non-zero and the measured values of critical exponent $\beta$ are very close to $1/2$, which indicates the pion condensation is of second order and belong to 4D mean field class. To further confirm the order of this phase transition, more study about the relaxation phenomenon may be helpful.
However, in order to got a critical exponent beyond mean field theory, on one hand, the large-$N_c$ correction should be take into consideration~\cite{D.Hou, D.Hou1}, on the other hand, a full back-reaction model including the interaction fo gluodynamics and chiral dynamics should be considered in a more realistic holographic model.
The relationship between $\pi_1$ and $\sigma$ indicate that the absolute chiral condensation, $\widetilde{\sigma}=\sqrt{\sigma^2+\pi_1^2}$ cannot be enhanced by the pion condensation at high temperature, however it is first enhanced by pion condensation, then steeply decreases to zero along the increasing of $\mu_I$ at low temperature region. When temperature infinitely approaches to zero, the enhancement will approach to infinite large pion condensation region, which coincide with the zero temperature holography hard wall results in Ref~\cite{enhancement1}. It is because when temperature tends to zero, the dilaton term will approach to one as $z_h$ approach to zero, so that the soft wall boundary will back to the hard wall cut.

After studying the condensation in detail, we get the QCD phase diagram on the $\mu_I-T$ plane.
In which the non-zero pion condensation possesses a convex shape area;  the zero temperature critical point $\mu_{I, T=0}^c\approx m_\pi$ is  extrapolated from the low temperature tendency, which well coincides with the results in Refs.~\cite{B.Brandt,D.T.Son,enhancement1};
the chiral crossover boundary interposes the pion condensation region at the tricritical point, as chiral crossover can be seen as a signal of color deconfinement transition, the chiral crossover boundary in the pion condensation region indicates a crossover from BEC to BCS. In addition, the coincidence between the boundary EOMs and the momentum space EOMs at $\omega=0$ and $\mu_I=\mu_I^c$, proves the pion mass $m_{\pi_1}(\mu_I^c,T_c)=0$ on the pion condensation phase boundary.

\section*{Acknowledgement}
H.L. is supported by the National Natural Science Foundation of
China under Grant No. 11405074. D.L. is supported by the National Natural Science Foundation of China under Grant No.11805084, the PhD Start-up Fund of Natural Science Foundation of Guangdong Province under Grant No. 2018030310457 (and  Guangdong  Pearl River Talents Plan under  Grant \quad No. 2017GC010480). 
\end{multicols}

\vspace{10mm}

\begin{multicols}{2}

\end{multicols}

\vspace{-1mm}
\centerline{\rule{80mm}{0.1pt}}
\vspace{2mm}

\begin{multicols}{2}

\end{multicols}
\clearpage
\end{document}